\documentstyle[12pt]{article}
\begin{document}
\begin{titlepage}

\begin{flushright}
{\Large TUIMP-TH-95/76}
\end{flushright}
\vspace{0.2cm}
\begin{center}
{\Large \bf Supersymmetric QCD Corrections to Top Quark Pair Production in 
Photon-Photon Collision}
\footnote{Work supported by the National Natural Science Foundation of
China, the Fundamental Research Foundation of Tsinghua University, and
the State Commission of Science and Technology of China}
\vspace{.5in}

{\bf Hua Wang $^b$,~Chong-Sheng Li $^{a,c}$,~Hong-Yi Zhou $^{a,b}$,~
Yu-Ping Kuang $^{a,b}$}\\
\vspace{.3in}
    
{\small
  $^a$ CCAST (World Laboratory), \hspace{.2cm}
      P. O.\hspace{0.2cm}  Box 8730, Beijing 100080, P.R. China,\\
	 
  $^b$ Institute of Modern Physics and Department of Physics,\\
   Tsinghua University, Beijing 100084, P.R. China \\	 

  $^c$ Physics Department, Peking University, Beijing 100871, P.R. China}\\

\vspace{.5in}

\end{center}

\begin{footnotesize}
\begin{center}\begin{minipage}{5in}

\begin{center} ABSTRACT\end{center}

Supersymmetric QCD corrections to top quark pair 
production by $\gamma \gamma$ fusion are calculated in the 
minimal supersymmetric standard model taking into account the effects of 
stops in the corrections to the total cross-section of $t \bar{t}$ production 
at the future $e^+e^- $ linear collider. We find that the relative correction 
can be a few percent for reasonable values of the parameters.

\end{minipage}\end{center}
\end{footnotesize}
\vspace{.7in}

~~PACS number: 14.80Dq; 12.38Bx; 14.80.Gt

\vspace{0.5in}

\vfill
\end{titlepage}

\eject
\rm
\baselineskip=0.36in

\begin{flushleft} {\bf 1. Introduction} \end{flushleft}

 Recently, the top quark was found experimentally by the CDF and D0 
Collaborations at Fermilab\cite{CDFD0}. The measured top-quark mass, 
$~176\pm 10^{+13}_{-12}~\rm GeV$, is close to the central value of that 
obtained from the best fit of the standard model (SM) to the latest LEP data. 
This is a remarkable success of the SM. However, there are a number of unsolved
theoretical puzzles in the SM, and the latest LEP data on the
 branching ratio $R_b$ of $~Z\rightarrow b{\bar b}~$ deviates from the SM
prediction by 3.7 standard deviation\cite{Hagiwara}. These lead to 
more interest in considering possible new physics beyond the SM. Processes
with top quarks may be good for testing new physics since the top quark is the
heaviest particle yet found. Among various models of new physics so far
considered, supersymmetry (SUSY) is a promising one at present.
The simplest and interesting SUSY model is the minimal supersymmetric 
extension of the standard model (MSSM) \cite{MSSM}. For solving the gauge 
hierarchy problem, SUSY should be broken at energies around $1~\rm TeV$, and 
thus SUSY particles in the MSSM may be within the reach of future
colliders. At the future $e^+e^-$ linear collider with 
center-of-mass energy  $500~\rm GeV$ to $1.5~\rm TeV$, the $e^+e^- \rightarrow 
t\bar{t}$ event rate would be around $10^4$/yr, comparable with the Tevatron,
however, the events would be easier to extract. It is possible to separately 
measure all of the various production and decay form factors of the top quark
at the level of a few percent \cite{LI5}. Thus theoretical calculations of
the radiative corrections to the production and decay of the top quark is of 
importance. SUSY corrections to $t\bar{t}$ pair production in $e^+e^-$ 
annihilation has been calculated in Ref.\cite{CHANG}. At the $e^+e^-$ 
linear collider, hard photons can be obtained by laser backscattering. The 
intense $\gamma $ beams are generated by backward Compton scattering of soft 
photons from a laser of a few eV energy \cite{OJP10}. The luminosity 
distribution over the $\gamma \gamma $ invariant mass is broad and contains 
an abundant number of very energric photons. The hard photon beam has 
approximately the same luminosity as the original electron beam.
Therefore photon collisions are also good processes for testing new physics.

In this paper we investigate the SUSY QCD correction to the top quark 
production by the process $\gamma \gamma \rightarrow t{\bar t}$ in MSSM model.
In Sec. II, we give our calculation of the SUSY QCD corrections to the
scattering cross-section. The numerical results of the cross-section  
are given in Sec III.  In recent years there have been renewed 
interest in the possibility of very light gluinos, with mass 
$m_{\tilde{g}} \leq 5~\rm GeV$ \cite{ADJ1}. Our conclusion is that, with
such light gluinos, the corrections can be large enough to be
experimentally testable. 

\begin{flushleft} {\bf 2. The Cross-Section} \end{flushleft}

The total cross section of the production of $t\bar{t}$ in $\gamma \gamma $ 
collisions at the $e^+e^-$ collider is obtained by folding the 
elementary cross-section for the processes $\gamma \gamma \rightarrow 
t\bar{t}$ with the photon luminosity \cite{OJP}
\begin{equation}                                      
 \sigma _s = \int_{2m_t/ \sqrt{\hat{s}}}^{x_{max}}dz\frac{dL_{\gamma \gamma }}
{dz} \sigma (\gamma \gamma \rightarrow t \bar{t}~ at~ \hat{s} = z^2 s) ,
\end{equation}
where $ \sqrt{s} (\sqrt{\hat{s}}) $ is the $e^+e^-(\gamma \gamma ) $ 
center-of-mass energy and the quantity $ \frac{dL_{\gamma \gamma }}{dz} $ is 
the photon luminosity defined as \cite{OJP} 
\begin{equation}                                      
 \frac{dL_{\gamma \gamma }}{dz}=2z \int_{\frac{z^2}{x_{max}}}^{x_{max}}
 \frac{dx}{x} F_{\gamma /e}(x)F_{\gamma /e}(z^2/x) .
\end{equation}
For unpolarized initial electrons and laser,the energy spectrum of the
back-scattered photon is given by \cite{OJP}
\begin{eqnarray}                                      
 F_{\gamma /e} & = & \frac{1}{D(\xi )}[1-x+\frac{1}{1-x}-\frac{4x}{\xi (1-x)}
 +\frac{4x^2}{\xi ^2 (1-x)^2}] \\
 D(\xi ) & = & (1-\frac{4}{\xi }-\frac{8}{ \xi ^2})ln(1+\xi )+\frac{1}{2}
 +\frac{8}{\xi }-\frac{1}{2(1+\xi )^2} ,
\end{eqnarray} 
where $ \xi = 4E_0 \omega _0/m_e^2 $ with $ m_e $ and $ E_0 $ the incident 
electron mass and energy, respectively, and $ \omega _0 $  the laser-photon
energy, x is the fraction of energy of the incident electron carried by 
the back-scattered photon. Following Ref.\cite{OJP}, we choose $\xi $ and 
$x_{max}$ to be
\begin{equation}                                     
 \xi =2(1+\sqrt{2}) \approx 4.8,~~~x_{max} \approx 0.83,~~~D(\xi ) 
 \approx 1.8 .
\end{equation}
The Feynman diagrams contributing to the SUSY O($\alpha _s \alpha _e$)
corrections are shown in Fig.1. In our calculation, we use dimensional 
regularization 
and take the on-shell renormalization scheme.

In Fig.1 one has to include the contributions of both stops. As is well-known 
\cite{ADJ12}, the supersymmetric partner of left- and right-handed massive 
quarks mix with each other. The mass eigenstates $\tilde{q}_1$ and 
$\tilde{q}_2$ are related to the current eigenstates $\tilde{q}_L$ and 
$\tilde{q}_R$ by
\begin{equation}                                       
  \tilde{q} _1 = \tilde{q} _L cos \theta _q + \tilde{q} _R sin \theta _q,
~~~\tilde{q} _2 = -\tilde{q} _L sin \theta _q + \tilde{q} _R cos \theta _q .
\end{equation}
The mixing angle $\theta _q$ as well as the masses $m_{\tilde{t}_1},
m_{\tilde{t}_2}$ of the physical stops can be calculated from the following
mass matrices \cite{ADJ}
 \begin{equation}                                      
  M^2_t =\left(
	   \begin{array} {ll}
	     m^2_{\tilde{t} _L} + m^2_t +0.35D_Z & -m_t(A_t+\mu cot \beta) \\
	     -m_t(A_t+\mu cot \beta) & m^2_{\tilde{t}_R} + m^2_t + 0.16D_Z 
	   \end{array}
	 \right),
\end{equation}
where $ D_Z=M_Z^2cos 2 \beta$, $tan \beta $ is the ratio of the vacuum 
expectation values of the two neutral Higgs fields of the MSSM,
$ m_{\tilde{t} _L},m_{\tilde{t} _R} $ are soft breaking masses, 
$ A_t $ are parameters describing the strength of nonsupersymmetric trilinear
scalar interactions, and $\mu$ is the supersymmetric Higgs(ino) mass
which also appears in the trilinear scalar vertices. 
 
In the presence of squark mixing, the squark-quark-gluino interaction 
Lagrangian is given by
\begin{equation}                                        
\begin{array} {lll}
 L_{\tilde{g} \tilde{q} \bar{q}} & = & -i \sqrt{2} g_s T^a \bar{q} 
 [(cos \theta _q \tilde{q} _1 - sin \theta _q \tilde{q} _2) 
 \frac{1+ \gamma _5}{2} \\
 & & -(sin \theta _{q} \tilde{q} _1 + cos \theta _q \tilde{q} _2)
 \frac{1- \gamma _5}{2}]\tilde{g} _a + h.c. ,
\end{array}
\end{equation}
where $g_s$ is the strong coupling constant and $T^a$ are $SU(3)_C$ generators.
 
To the precision of the O($\alpha _s \alpha _e$) SUSY corrections, the 
renormalized amplitude for $\gamma \gamma \rightarrow t\bar{t} $ is 
\begin{equation}                                         
  M_{ren} = M_0+\delta M^{self}+\delta M^{vertex}+\delta M^{box}+\delta M^s ,
\end{equation}  
where $M_0$ is the tree-level amplitude and $\delta M$ represents SUSY
QCD correctrions, which are
\begin{equation}                                          
M_0 ~ = ~ \epsilon ^{\mu }(p_4)\epsilon ^{\nu }(p_3) \bar{u}(p_2) ,
T_{\mu \nu }v(p1) \\
\end{equation}
with
\begin{equation}                                           
  T_{\mu \nu } ~= ~ \frac{-i4 \pi \alpha _e}{t-m_t^2}\gamma _{\mu }
  (\rlap/p_3-\rlap/p_1+m_t)\gamma _{\nu }
 +\frac{-i4 \pi \alpha _e}{u-m_t^2}\gamma _{\nu }(\rlap/p_4-\rlap/p_1+m_t)
 \gamma _{\mu }, 
 \end{equation}                                          
 and
 \begin{eqnarray}                                      
 \delta M^{self} & = & \delta M^{self(t)}+\delta M^{self(u)} \\
 \delta M^{vertex} & = & \delta M^{vertex(t)}+\delta M^{vertex(u)} \\
 \delta M^{box} & = & \delta M^{box(t)}+\delta M^{box(u)} ,
\end{eqnarray}
\begin{equation}\begin{array}{lll}                          
 \delta M^{self(t)} & = & \epsilon ^{\mu }(p_4)\epsilon ^{\nu }(p_3) 
 \bar{u}(p_2)\frac{Q_t^2}{(t-m_t^2)^2}\gamma _{\mu} (\rlap/p_2-\rlap/p_4+m_t)
 (-i\hat{\Sigma}) \\
 & & \times (\rlap/p_3-\rlap/p_1+m_t)\gamma _{\nu}v(p_1) ,
\end{array}\end{equation}
\begin{equation}\begin{array}{lll}                          
 \delta M^{vertex(t)} & = & \epsilon ^{\mu }(p_4) \epsilon ^{\nu }(p_3) 
 \bar{u}(p2) \{ \frac{i}{t-m_t^2}[i\hat{\Lambda }_{\mu }^{(t1)}
 (\rlap/p_3-\rlap/p1+m_t) \\
 & & \times (-iQ_t^2 \gamma _{\nu } )+(-iQ_t^2 \gamma _{\mu })
 (\rlap/p_2-\rlap/p_4+m_t)i\hat{\Lambda }_{\nu }^{(t2)}]\}v(p_1) ,\\
\end{array}\end{equation}
\begin{equation}\begin{array}{lll}                          
 \delta M^{box(t)} & = &\epsilon ^{\mu }(p_4)\epsilon ^{\nu }(p_3) 
 \bar{u}(p_2)Q_t ^2 \{\gamma _{\nu } \gamma _{\mu }f^{box(t)}_1+\gamma _{\mu }
 \gamma _{\nu }f^{box(t)}_2 + p_{1 \nu }\gamma _{\mu }f^{box(t)}_3 \\
 & & +p_{1 \mu } \gamma _{\nu }f^{box(t)}_4 + p_{2 \nu }\gamma _{\mu }
 f^{box(t)}_5 + p_{2 \mu }\gamma _{\nu }f^{box(t)}_6 + p_{1 \mu }p_{1 \nu }
 f^{box(t)}_7 \\
 & & +p_{1 \mu }p_{2 \nu }f^{box(t)}_8 + p_{2 \mu}p_{1 \nu }f^{box(t)}_9 
 + p_{2 \mu }p_{2 \nu }f^{box(t)}_{10} + \rlap/p_4 \gamma _{\nu }
 \gamma _{\mu }f^{box(t)}_{11} \\
 & & +\rlap/p_4\gamma _{\mu }\gamma _{\nu }f^{box(t)}_{12} 
 + \rlap/p_4 p_{1 \nu }\gamma  _ {\mu }f^{box(t)}_{13} 
 + \rlap/p_4 p_{1 \mu } \gamma _{\nu }f^{box(t)}_{14} \\
 & & +\rlap/p_4 p_{2 \nu}\gamma _{\mu }f^{box(t)}_{15} 
 + \rlap/p_4 p_{2 \mu }\gamma _{\nu } f^{box(t)}_{16} 
 + \rlap/p_4 p_{1 \mu} p_{1 \nu }f^{box(t)}_{17} \\
 & & +\rlap/p_4 p_{1 \mu }p_{2 \nu }f^{box(t)}_{18} 
 + \rlap/p_4 p_{2 \mu }p_{1 \nu }f^{box(t)}_{19} 
 + \rlap/p_4 p_{2 \mu}p_{2 \nu }f^{box(t)}_{20} \}v(p_1),
\end{array}\end{equation}

\begin{eqnarray}                                                
\delta M^s = \epsilon _{\mu }(p_4) \epsilon ^{\mu }(p_3) \bar{u}(p_2)
Q_t^2Sv(p_1) ,
\end{eqnarray}
with
\begin{eqnarray}                                              
 -i \hat{\Sigma } & = & f^{\Sigma (t)}_1+\rlap/t f^{\Sigma (t)}_2 \\
 i \hat{\Lambda }_{\mu }^{(t1)} & = & \gamma _{\mu }
 f^{\Lambda _{\mu }^{(t1)}}_1 + p_{2 \mu }f^{\Lambda _{\mu }^{(t1)}}_2 
 + \rlap/p_4 p_{2 \mu }f^{\Lambda _{\mu }^{(t1)}}_5 \\
 i \hat{\Lambda }_{\nu }^{(t2)} & = & \gamma _{\nu }
 f^{\Lambda _{\nu }^{(t2)}}_1 + p_{1 \nu }f^{\Lambda _{\nu }^{(t2)}}_2 
 + \rlap/p_3 p_{1 \nu }f^{\Lambda _{\nu }^{(t2)}}_5 ,
\end{eqnarray}
where $Q_t=2/3$, S and the form-factors $f_i$ are given in the Appendix. 
Instead of calculating the square of the amplitudes explicitly, we calculate  
the amplitudes numerically by using the method of Ref. \cite{ZEPPEN}. 
This greatly simplifies our calculations.   

We only explicitly give the results for the t-channel contributions to 
the SUSY corrections. The u-channel results can be obtained by the 
following substitutions
\begin{equation}\begin{array}{l}                          
p_3\leftrightarrow p_4,\;\;
T^a\leftrightarrow T^b,\;\;
\hat{t}\leftrightarrow\hat{u}.
\end{array}\end{equation}

\vspace{1cm}
\begin{flushleft} {\bf 3. Nemerical Results and Conclusion} \end{flushleft}

Now we present the numerical results. We take $m_t=176~\rm GeV$, and use the 
two-loop running coupling constant $\alpha _s$ and the input 
$\alpha _e=1/128$. For the SUSY parameters involved in our 
calculations, we see that once $tan \beta $ and $m_{\tilde{t_L}}$ are fixed, 
we are, in general, free to choose two independent parameters in the stop mass 
matrix, namely $m_{\tilde{t_R}}$ and $A_t+\mu cot \beta$,
or equivlantly $m_{\tilde{t_R}}$ and $m_{\tilde{t}_1}$. To avoid the 
singularities at small angles, we take the following kinematical cuts
\begin{equation}\begin{array}{l}                          
|\eta|<2.5,\;\;p_T>20\;\rm GeV.
\end{array}\end{equation} 
In this kinematical region the relative corrections are actually large.
In the numerical calculation, we have checked our program with the requirement 
of gauge invariance to the accuracy $10^{-10}$.

In Figs.2-8 we give the numerical results in a simple case in which we
set $tan \beta =1$ and $m_{\tilde{t_L}}=m_{\tilde{t_R}}=m_{\tilde{q}}$ 
(coresponding to the mixing angle equal to $\pi /4$). We choose 
$m_{\tilde{t_1}}$ as the light stop mass and require it to be heavier than 
$45~\rm GeV$ \cite{ADJ}.

 Fig.2~(Fig.3) show the dependence of the 
corrections on $m_{\tilde{t_1}}$ for fixed $m_{\tilde{g}}=3~\rm GeV$, 
$m_{\tilde{q}}=150~\rm GeV~(450~\rm GeV)$ and $\sqrt{s}=0.5~\rm TeV
~(1.5~\rm TeV)$. We see that the corrections can be either positive or 
negative depending on the light stop mass. The corrections become their 
negative maxima at $m_{\tilde{t_1}}=170~\rm GeV$. Fig.4~(Fig.5) show the 
dependence of the corrections on $\sqrt{s}$ for fixed $m_{\tilde{g}}=
3~\rm GeV$, $m_{\tilde{q}}=450~\rm GeV$ and $m_{\tilde{t}_1}=50~\rm GeV~
(150\rm ~GeV)$. The corrections are positive when $m_{\tilde{t}_1}=
50~\rm GeV$, and negative when $m_{\tilde{t}_1}=150~\rm GeV$. 
When $\sqrt{s}$ varies from $0.5~\rm TeV$ to $1.5~\rm TeV$, the 
corrections vary from $2\%~(-3.6\%)$ to $0.9\%~(-3.8\%)$ in the case of
$m_{\tilde{t}_1}=50~\rm GeV~(150~\rm GeV)$. Fig.6 (Fig.7) shows the dependence 
of the corrections on $m_{\tilde{q}}$ for fixed $m_{\tilde{g}}=3~\rm GeV$,
$m_{\tilde{t}_1}=50~\rm GeV ~(150~\rm GeV)$ and $\sqrt{s}=0.5~\rm TeV
~(1.5~\rm TeV)$. The relative corrections increase (decrease) as 
$m_{\tilde{q}}$ varies from $50~\rm GeV$ to $450~\rm GeV$ in the case of
$m_{\tilde{t}_1}=50~\rm GeV ~(150~\rm GeV)$. The largest relative correction
in Fig.7 can exceed $-5\%$.
Fig.8 shows the dependence of the corrections on $m_{\tilde{g}}$ for fixed 
$m_{\tilde{t}_1}=150~\rm GeV$, $m_{\tilde{q}}=450~\rm GeV$ and
$\sqrt{s}=0.5~\rm TeV ~(1.5~\rm TeV)$. If the glunios are light\cite{ADJ1}, 
e.g. $m_{\tilde{g}}=3~\rm GeV$, 
the corrections can reach $-4\%$. Whereas for heavy glunios as 
$m_{\tilde{g}} \geq 100~\rm GeV $, the corrections are less than $1\%$.
   
We have also done the numerical calculations for $tan\beta =10$ and found that
the corrections are not sensitive to the value of $tan\beta$. For example, 
with $m_{\tilde{g}}=3~\rm GeV$, $m_{\tilde{t}_1}=150~\rm GeV$, 
$m_{\tilde{q}}=450~\rm GeV$
and $\sqrt{s}=0.5~\rm TeV$, we get $\Delta \sigma = -3.60$ for $tan\beta =1$ 
and $\Delta \sigma = -3.58$ for $tan\beta =10$.
 
We conclude that, in the case with stop mixing and light gluinos like
$m_{\tilde{g}}$ around a few $\rm GeV$, the SUSY QCD corrections
to the cross-section of top quark pair production in  $\gamma \gamma $ fusion 
at the $e^+e^-$ collider can be as large as a few percent of the tree-level 
cross-section. This is experimentally testable.
 
\newpage
\begin{center}
{\bf Appendix} 
\end{center}

We give here the form factors for the matrix element.  They are 
written in terms of the usual one-, two-, three- and four -point scalar 
loop integrals of Ref.\cite{VELTMAN}.  
\begin{eqnarray} 
  F_i & = & m_{\tilde{g}}(a_i^2-b_i^2),~~ G_i = a_i^2+b_i^2 \\
 a_1=-b_2 & = & \frac{1}{\sqrt{2}}(cos\theta - sin\theta ),~~a_2=b_1 = 
\frac{1}{\sqrt{2}}(cos\theta + sin\theta )
\end{eqnarray}
\begin{eqnarray}
 f_1^{\Sigma (t)} & = & i \frac{4 \alpha _s \alpha _e}{3} \sum\limits_{i}
 [(F_iB_0 - m_t ZV_i - DM_i)] \\
 f_2^{\Sigma (t)} & = & i \frac{4 \alpha _s \alpha _e}{3} \sum\limits_{i}
 [G_i(B_0+B_1) + ZV_i ] ,
\end{eqnarray}
where $B_0,B_1(t,m_{\tilde{t_i}},m_{\tilde{g}})$ are  2-point Feynman 
integrals \cite{VELTMAN}.
\begin{eqnarray}
 f_1^{\hat{\Lambda }_{\mu }^{(t1)}} & = & -i \frac{4 \alpha _s \alpha _e}{3} 
 \sum\limits_{i} [2G_iC_{24}+ZV_i] \\
 f_2^{\hat{\Lambda }_{\mu }^{(t1)}} & = & -i \frac{4 \alpha _s \alpha _e}{3} 
 \sum\limits_{i} [2G_im_t(C_{11}+C_{21})-2F_i(C_0-C_{11})] \\
 f_5^{\hat{\Lambda }_{\mu }^{(t1)}} & = & -i \frac{4 \alpha _s \alpha _e}{3} 
 \sum\limits_{i} [2G_i(-C_{12}-C_{23})] ,
\end{eqnarray}
where $C_0,C_{ij}(-p_2,p_4,m_{\tilde{g}},m_{\tilde{t_i}},m_{\tilde{t_i}})$ 
are 3-point Feynman integrals \cite{VELTMAN}.
\begin{eqnarray}
 f_1^{\hat{\Lambda }_{\nu }^{(t2)}} & = & -i \frac{4 \alpha _s \alpha _e}{3} 
 \sum\limits_{i} [2G_iC_{24}+ZV_i] \\ 
 f_2^{\hat{\Lambda }_{\nu }^{(t2)}} & = & -i \frac{4 \alpha _s \alpha _e}{3} 
 \sum\limits_{i} [2G_im_t(-C_{11}-C{21})+2F_i(C_0+C_{11})]\\ 
 f_5^{\hat{\Lambda }_{\nu }^{(t2)}} & = & -i \frac{4 \alpha _s \alpha _e}{3} 
 \sum\limits_{i} [-2G_i(C_{12}+C_{23})],
\end{eqnarray}
where $C_0,C_{ij}(p_1,-p_3,m_{\tilde{g}},m_{\tilde{t_i}},m_{\tilde{t_i}})$ 
are the 3-point Feynman integrals \cite{VELTMAN}. 
\begin{eqnarray}
 f_1^{box(t)} & = & i \frac{4 \alpha _s \alpha _e}{3} \sum\limits_{i} 
 [2F_iD_{27}-2G_im_tD_{311}] \\
 f_2^{box(t)} & = & i \frac{4 \alpha _s \alpha _e}{3} \sum\limits_{i} 
 [2F_iD_{27}-2G_im_tD_{311}] \\
 f_3^{box(t)} & = & i \frac{4 \alpha _s \alpha _e}{3} \sum\limits_{i} 
 [4G_i(D_{27}+D_{312})] \\
 f_4^{box(t)} & = & i \frac{4 \alpha _s \alpha _e}{3} \sum\limits_{i} 
 [4G_iD_{313}] \\
 f_5^{box(t)} & = & i \frac{4 \alpha _s \alpha _e}{3} \sum\limits_{i} 
 [4G_i(-D_{311}+D_{312})] \\
 f_6^{box(t)} & = & i \frac{4 \alpha _s \alpha _e}{3} \sum\limits_{i} 
 [4G_i(-D_{27}-D_{311}+D_{313})] \\
 f_7^{box(t)} & = & i \frac{4 \alpha _s \alpha _e}{3} \sum\limits_{i} 
 [4F_i(D_{13}+D_{26})-4G_im_t(D_{25}+D_{310})] \\
 f_8^{box(t)} & = & i \frac{4 \alpha _s \alpha _e}{3} \sum\limits_{i} 
 [4F_i(-D_{25}+D_{26})+4G_im_t(D_{35}-D_{310})] ,
\end{eqnarray}
\begin{equation}\begin{array}{lll} 
f_9^{box(t)} & = & i \frac{4 \alpha _s \alpha _e}{3}\sum\limits_{i} 
[4F_i(-D_0-D_{11}+D_{13}-D_{12}-D_{24}+D_{26}) + 4G_i \times  \\
 &  & m_t(D_{11}+D_{21}-D_{25}+D_{24}+D_{34}-D_{310})],
\end{array}\end{equation}
\begin{equation}\begin{array}{lll}
 f_{10}^{box(t)} & = & i \frac{4 \alpha _s \alpha _e}{3} \sum\limits_{i} 
 [4F_i(D_{11}-D_{12}+D_{21}-D_{24}-D_{25}+D_{26}) +4G_i \times \\
 &  & m_t(-D_{21}+D_{24}-D_{31}+D_{34}+D_{35}-D_{310})],
\end{array}\end{equation}
\begin{eqnarray}
 f_{11}^{box(t)} & = & i \frac{4 \alpha _s \alpha _e}{3} \sum\limits_{i} 
 [2G_i(D_{312}-D_{313})] \\
 f_{12}^{box(t)} & = & i \frac{4 \alpha _s \alpha _e}{3} \sum\limits_{i} 
 [2G_i(D_{312}-D_{313})] \\
 f_{13}^{box(t)} & = & 0 \\
 f_{14}^{box(t)} & = & 0 \\
 f_{15}^{box(t)} & = & 0 \\
 f_{16}^{box(t)} & = & 0 \\
 f_{17}^{box(t)} & = & i \frac{4 \alpha _s \alpha _e}{3}\sum\limits_{i} 
 [4G_i(-D_{23}+D_{26}+D_{38}-D_{39})] \\
 f_{18}^{box(t)} & = & i \frac{4 \alpha _s \alpha _e}{3} \sum\limits_{i} 
 [4G_i(D_{37}+D_{38}-D_{39}-D_{310})] ,
\end{eqnarray}
\begin{equation}\begin{array}{lll}
 f_{19}^{box(t)} & = & i \frac{4 \alpha _s \alpha _e}{3} \sum\limits_{i} 
 [4G_i(D_{13}-D_{12}-D_{23}-D_{24}+D_{25} \\
 & & +2D_{26}-D_{22}-D_{36}+D_{38}-D_{39}+D_{310})] ,
\end{array}\end{equation}
\begin{equation}\begin{array}{lll}
 f_{20}^{box(t)} & = & i \frac{4 \alpha _s \alpha _e}{3} \sum\limits_{i} 
 [4G_i(-D_{22}+D_{24}-D_{25}+D_{26}+D_{34} \\
 & & -D_{35}-D_{36}+D_{37}+D_{38}-D_{39})] ,
\end{array}\end{equation}
\begin{eqnarray}
 S & = & i\frac{8}{3}\alpha _s \alpha _e \sum\limits_ {i} 
 [G_im_tC_{11}-F_iC_0](-p_2,p_4+p_3,m_{\tilde{g}},m_{\tilde{t_i}},
 m_{\tilde{t_i}}) ,
\end{eqnarray}
where $D_0,D_{ij},D_{ijk}(-p_2,p_4,p_3,m_{\tilde{g}},m_{\tilde{t_i}},
m_{\tilde{t_i}},m_{\tilde{t_i}}) $ are 4-point Feynman integrals\cite{VELTMAN}.

The renormalization constants are
\begin{equation}
\begin{array}{lll} 
ZV_i & = & -G_i[B_0+B_1](p,m_{\tilde{t_i}},m_{\tilde{g}})|_{p^2=m_t^2} 
-[2m^2_tG_i\frac{\partial^{2}}{\partial p^{2}}(B_0+B_1) \\
& & -2m_tF_i \frac{\partial ^{2}}{\partial p^2}B_0](p,m_{\tilde{t_i}},
m_{\tilde{g}})|_{p^2=m^2_t} ,
\end{array}\end{equation}
\begin{equation}
DM_i = [F_iB_0+m_tG_i(B_0+B_1)](p,m_{\tilde{t_i}},m_{\tilde{g}})|_{p^2=m^2_t} .
\end{equation}

\newpage

\begin{center}
{\large \bf Figure Captions}
\end{center}

\parindent=0pt

{\bf Fig.1}  Tree-level Feynaman diagrams and $O(\alpha _s \alpha _e)$
SUSY QCD corrections to $\gamma \gamma \rightarrow t\bar t $.

{\bf Fig.2}  Relative SUSY QCD corrections to $\gamma \gamma $ fusion cross-section  
versus $m_{\tilde{t}_1}$ for $m_{\tilde{g}}=
3~\rm GeV,m_{\tilde{q}}=150~\rm GeV $.

{\bf Fig.3} Same as fig.2,but for $m_{\tilde{g}}=3~\rm GeV,
m_{\tilde{q}}=450~\rm GeV $.

{\bf Fig.4} Same as Fig.2, but versus $\sqrt{s}$ for 
$m_{\tilde{g}}=3~\rm GeV,m_{\tilde{t}_1}=50~\rm GeV,
m_{\tilde{q}}=450~\rm GeV $.

{\bf Fig.5} Same as Fig.2, but versus $\sqrt{s}$ for 
$m_{\tilde{g}}=3~\rm GeV,m_{\tilde{t}_1}=150~\rm GeV,
m_{\tilde{q}}=450~\rm GeV $.
  
{\bf Fig.6} Same as Fig.2, but versus $m_{\tilde{q}}$ for 
$m_{\tilde{g}}=3~\rm GeV,m_{\tilde{t}_1}=50~\rm GeV $.

{\bf Fig.7} Same as Fig.2, but versus $m_{\tilde{q}}$ for 
$m_{\tilde{g}}=3~\rm GeV,m_{\tilde{t}_1}=150~\rm GeV $.

{\bf Fig.8} Same as Fig.2, but versus $m_{\tilde{g}}$ for 
$m_{\tilde{q}}=450~\rm GeV,m_{\tilde{t}_1}=150~\rm GeV $.

\end{document}